\documentclass[prl, reprint, amsmath, amssymb, amsfonts,superscriptaddress,nofootinbib]{revtex4-2}

\usepackage[utf8]{inputenc}
\usepackage{color}
\usepackage{braket}
\usepackage{graphicx}

\usepackage[caption=false]{subfig}

\begin{document}

\title{Accelerated cosmological expansion from pseudo-Hermiticity}
\author{Edmund J.~Copeland}
\email{ed.copeland@nottingham.ac.uk}
\affiliation{School of Physics and Astronomy, University of Nottingham,\\ University Park, Nottingham NG7 2RD, United Kingdom}
\author{Andrei Lazanu}
\email{andrei.lazanu@manchester.ac.uk}
\affiliation{Department of Physics and Astronomy, University of Manchester,\\ Oxford Road, Manchester M13 9PL, United Kingdom}
\author{Peter Millington}
\email{peter.millington@manchester.ac.uk}
\affiliation{Department of Physics and Astronomy, University of Manchester,\\ Oxford Road, Manchester M13 9PL, United Kingdom}
\author{Esra Sablevice}
\email{esra.sablevice@manchester.ac.uk}
\affiliation{Department of Physics and Astronomy, University of Manchester,\\ Oxford Road, Manchester M13 9PL, United Kingdom}

\date{23 July 2025}

\begin{abstract}
    
   We show that a well-studied pseudo-Hermitian field theory composed of two complex scalar fields can generate accelerated cosmological expansion through a novel mechanism. The dynamics is unique to the pseudo-Hermitian field theory, and it arises in the regime of broken antilinear symmetry, wherein a growth instability from the resulting complex eigenspectrum competes with the Hubble damping. The azimuthal components of the complex scalar fields asymptote to a constant rate of rolling at late times, reminiscent of motion around the infinite staircase of M.~C.~Escher's lithograph ``Ascending and Descending''. The resulting centripetal acceleration drives the radial components of the field away from the minimum of the potential, and the system generates a self-sustaining and constant Hubble rate at late times, even when tuning the minimum of the potential such that the classical vacuum energy is vanishing. This result evidences the potential to generate novel and physically relevant dynamics that are unique to pseudo-Hermitian field theories, and that their regimes of broken antilinear symmetry can be physically relevant in dynamical spacetimes.
    
\end{abstract}

\maketitle

Since the seminal works by Bender and Boettcher~\cite{Bender:1998ke}, and Mostafazadeh~\cite{Mostafazadeh:2001jk}, there has been an explosion of interest in quantum theories with non-Hermitian Hamiltonians.  The viability of these theories, in terms of the reality of their eigenspectra~\cite{Bender:1998ke} and the existence of a unitary evolution~\cite{Bender:2002vv}, is ensured by the existence of an unbroken antilinear symmetry of the Hamiltonian other than Hermitian conjugation~\cite{Mannheim:2015hto}, in many cases symmetry under the combined action of parity ($\mathcal{P}$) and time-reversal ($\mathcal{T}$).  This $\mathcal{P}\mathcal{T}$-symmetric quantum mechanics (and more generally pseudo-Hermitian quantum mechanics~\cite{Mostafazadeh:2001jk}) has found applications in various areas of physics, and, more recently, in the context of quantum field theories, where it may provide novel avenues for building extensions of the Standard Model of particle physics~\cite{Millington:2021fih} and, as we will show in this letter, cosmology.

Non-Hermitian matrices exhibit so-called exceptional points, where eigenvalues merge and the matrix becomes defective. These exceptional points represent the phase boundaries between regimes of unbroken and broken $\mathcal{P}\mathcal{T}$ symmetry.  When the $\mathcal{P}\mathcal{T}$ symmetry is unbroken, the eigenspectrum is real; when the $\mathcal{P}\mathcal{T}$ symmetry is broken, the eigenvalues come in complex-conjugate pairs.

In this letter, we consider the $c$-number field theory of two complex scalar 
fields $\phi_a$ ($a=1,2$) with mixing whose squared mass matrix can be non-Hermitian~\cite{Endnote1}:
\begin{align}
    \label{eq:Lag}
    -\mathcal{L}=&-\partial_{\nu}\tilde{\phi}_a^{\dag}\partial^{\nu}\phi_a+m_1^2\tilde{\phi}_1^{\dag}\phi_1-m_2^2\tilde{\phi}_2^{\dag}\phi_2\nonumber\\&-\mu^2\left(\tilde{\phi}_1^{\dag}\phi_2-\epsilon\tilde{\phi}_2^{\dag}\phi_1\right)-\frac{\lambda}{4}\left(\tilde{\phi}_1^{\dag}\phi_1\right)^2~\,,
\end{align}
where $m_1^2,m_2^2,\mu^2,\lambda>0$. The parameter  $\epsilon$ determines whether the Lagrangian is Hermitian ($\epsilon=-1$) or non-Hermitian ($\epsilon=+1$). Note that we use the mostly plus signature convention $(-,+,+,+)$, and there are two details warranting comment:

First, the conjugate field $\tilde{\phi}^{\dag}$ is denoted with a tilde because it need not be the usual Hermitian conjugate $\phi^{\dag}$ of $\phi$ in the non-Hermitian case ($\epsilon = +1$)~\cite{Endnote2}. Instead, the definition of the conjugate field involves a parity transformation (see Refs.~\cite{Alexandre:2020gah, Sablevice:2023odu}):
\begin{equation}
    \mathcal{P}:\ \phi(x)\longmapsto \phi^{\mathcal{P}}_a(x_{\mathcal{P}})=P_{ab}\tilde{\phi}_b(x)\;,
\end{equation}
where $x_{\mathcal{P}}$ is the parity-transformed coordinate, and the matrix $P={\rm diag}(1,-1)$ (with the overall sign a matter of convention), such that one field ($\phi_1$) transforms as a scalar and the other ($\phi_2$) as a pseudo-scalar. It is then readily confirmed that the non-Hermitian $c$-number Lagrangian~\eqref{eq:Lag} is $\mathcal{PT}$ symmetric, given
\begin{equation}
    \mathcal{T}:\ \phi_a(x)\longmapsto\phi_a^{\mathcal{T}}(x_{\mathcal{T}})=\phi_a^*(x)\,,
\end{equation}
where $x_{\mathcal{T}}$ is the time-reversed coordinate.

Second, the overall sign of the Lagrangian has been chosen \emph{a posteriori} to ensure positivity of the right-hand side of the Friedmann equation and a real Hubble parameter in what follows. As we will see, the kinetic terms are not necessarily positive for pseudo-Hermitian theories, and they need not be for the theory to be viable. It is well known, for example, that theories containing ghosts can be cured of instabilities when realised within the paradigm of $\mathcal{PT}$-symmetric quantum theory. Such is the case for the Lee~\cite{Bender:2004sv} and Lee--Wick models~\cite{Shalaby:2009re}, as well as the fourth-order quantum-mechanical Pais--Uhlenbeck oscillator~\cite{Bender:2007wu}.

The classical equations of motion, obtained by extremising the action corresponding to Eq.~\eqref{eq:Lag} with respect to variations in $\phi_a$ and $\tilde{\phi}_a^{\dag}$, are
\begin{subequations}
\begin{gather}
    \Box\phi_1+m_1^2\phi_1-\mu^2\phi_2-\frac{\lambda}{2}\tilde{\phi}_1^{\dag}\phi_1^2=0\,,\\
    \Box\phi_2-m_2^2\phi_2+\epsilon\mu^2\phi_1=0\,,\allowdisplaybreaks\\\Box\tilde{\phi}^{\dag}_1+m_1^2\tilde{\phi}^{\dag}_1+\epsilon\mu^2\tilde{\phi}^{\dag}_2-\frac{\lambda}{2}\tilde{\phi}_1^{\dag2}\phi_1=0\,,\\
    \Box\tilde{\phi}^{\dag}_2-m_2^2\tilde{\phi}_2^{\dag}-\mu^2\tilde{\phi}_1^{\dag}=0\,,
\end{gather}
\end{subequations}
with $\Box=\partial_{\nu}\partial^{\nu}$. It is clear from this system that, in the non-Hermitian case $\epsilon=+1$, the fields $\phi_a$ and $\tilde{\phi}_a^{\dag}$ are not Hermitian conjugates of one another.

This model exhibits spontaneous symmetry breaking, and the classical minima lie at
\begin{subequations}
\begin{gather}
    \phi_1\to v_1=\sqrt{\frac{2}{\lambda}\frac{m_1^2m_2^2-\epsilon\mu^4}{m_2^2}},\qquad \tilde{\phi}_1^{\dag}\to v_1,\\
    \phi_2 \to v_2=\frac{\epsilon \mu^2}{m_2^2}v_1,\qquad \tilde{\phi}_2^{\dag}\to -\epsilon v_2~,
\end{gather}
\end{subequations}
wherein we have set to unity an overall complex phase. Assuming $m_1^2m_2^2>\epsilon \mu^4$ such that $v_1,v_2\in\mathbb{R}$, fluctuations around this background satisfy
\begin{subequations}
\begin{gather}
    \Box\delta \phi_1-(\lambda v_1^2-m_1^2)\delta\phi_1-\mu^2\delta\phi_2-\frac{\lambda}{2}v_1^2\delta\tilde{\phi}^{\dag}_1=0\,,\\
    \Box\delta \phi_2-m_2^2\delta\phi_2+\epsilon \mu^2\delta\phi_1=0\,,\\
    \Box\delta \tilde{\phi}^{\dag}_1-(\lambda v_1^2-m_1^2)\delta\tilde{\phi}^{\dag}_1+\epsilon \mu^2\delta\tilde{\phi}^{\dag}_2-\frac{\lambda}{2}v_1^2\delta\phi_1=0\,,\\
    \Box\delta \tilde{\phi}^{\dag}_2-m_2^2\delta\tilde{\phi}_2^{\dag}- \mu^2\delta\tilde{\phi}^{\dag}_1=0\,,
\end{gather}
\end{subequations}
and the eigenmasses of the fluctuations are
\begin{subequations}
\begin{align}
    M_G^2&=0\,, \qquad
    M_0^2=m_2^2-\frac{\epsilon\mu^4}{m_2^2}\,,\\
    M_{\pm}^2&=\frac{m_2^2}{2}\Bigg[1+2\frac{m_1^2}{m_2^2}-3\frac{\epsilon\mu^4}{m_2^4}\nonumber\\&\phantom{=}\pm\sqrt{\left(1-2\frac{m_1^2}{m_2^2}\right)^2+2\left(1-6\frac{m_1^2}{m_2^2}\right)\frac{\epsilon\mu^4}{m_2^4}+9\frac{\epsilon^2\mu^8}{m_2^8}}\;\Bigg]\,,
\end{align}
\end{subequations}
wherein we see the expected massless Goldstone due to the spontaneous breaking of the model's global $U(1)$ symmetry. We also see that the squared eigenmass $M_0^2$ vanishes in the non-Hermitian case, merging with the Goldstone mode, when $\mu^2/m_2^2=1$. Above this exceptional point the squared mass $M_0^2$ becomes tachyonic.

The $\mathcal{PT}$-broken regime is often considered unphysical due to the presence of complex eigenmodes. This conclusion may be valid in Minkowski spacetime.  However, as we will now show, this need not be the case in dynamical spacetimes, wherein the presence of complex eigenmodes is already expected. For example, the time dependence of a homogeneous Klein--Gordon field in the de Sitter phase of the Friedmann--Lema\^{i}tre--Robertson--Walker (FLRW) spacetime evolves as $e^{-i(-3 i H)t}=a^{-3}$ due to the Hubble damping [see Eq.~\eqref{eq:eigenvalues} later].  It is this observation that we will now exploit to arrive at a novel mechanism for driving accelerated cosmological expansion; one that exploits the $\mathcal{PT}$-broken regime and the interplay between the Hubble damping and growth instability that arises from the non-Hermitian nature of the scalar sector.

Turning then to cosmology and the FLRW spacetime, the Lagrangian~\eqref{eq:Lag} is promoted to
\begin{equation}
    -\mathcal{L}=-g^{\nu\rho}\partial_{\nu}\tilde{\phi}^\dag_a\partial_{\rho}\phi_a-V(\tilde{\phi}^\dag_a,\phi_a)\,,
\end{equation}
where the potential $V(\tilde{\phi}^{\dag}_a,\phi_a)$ contains the non-derivative terms from Eq.~\eqref{eq:Lag}. The FLRW line element ${\rm d}s^2=-{\rm d}t^2+a^2(t)({\rm d}r^2+r^2{\rm d}\Omega^2_2)$ is not invariant under the Minkowski spacetime $\mathcal{PT}$ transformation, i.e., $t\longmapsto -t$ and $x^i\longmapsto -x^i$, since $a(t)=e^{Ht}$ in the de Sitter phase, and the antilinear symmetry of the theory is obfuscated.
However, moving to conformal time ${\rm d}\tau={\rm d}t/a(t)$, wherein $a(\tau)=-1/(H\tau)$, the line element becomes ${\rm d}s^2=a^2(\tau)(-{\rm d}\tau^2+{\rm d}r^2+r^2{\rm d}\Omega^2_2)$. This is invariant under the actions of the time-reversal operation $\tau\longmapsto-\tau$ and parity $x^i\longmapsto -x^i$, and both are symmetries of the flat slicing of the de Sitter phase of the spacetime. Moreover, for a static spacetime $a=\text{const.}$, the conformal time-reversal is just the regular time-reversal in Minkowski spacetime, since $t\propto \tau$ in this limit. Thus, the Lagrangian in FLRW has antilinear symmetry under the combined action of
\begin{subequations}
\begin{align}
    \mathcal{P}:&\ \phi_a(\tau, \vec{x}) \longmapsto \phi_a^{\mathcal{P}}(\tau,-\vec{x})=P_{ab}\tilde{\phi}_b(\tau,\vec{x})\,,\\
    \mathcal{T}:&\ \phi_a(\tau, \vec{x}) \longmapsto \phi_a^{\mathcal{T}}(-\tau,\vec{x})=\phi_a^*(\tau,\vec{x})\,.
\end{align}
\end{subequations}

Reverting back to cosmic time, and assuming spatially homogeneous field configurations, $\phi_a=\phi_a(t)$, the equations of motion on FLRW take the form
\begin{subequations}
\begin{align}
    -\ddot{\phi}_1-3H\dot{\phi}_1+m_1^2\phi_1-\mu^2\phi_2-\frac{\lambda}{2}\tilde{\phi}_1^{\dag}\phi_1^2=0\,,\\
    -\ddot{\phi}_2-3H\dot{\phi}_2-m_2^2\phi_2+\epsilon\mu^2\phi_1=0\,,\\
    -\ddot{\tilde{\phi}}^{\dag}_1-3H\dot{\tilde{\phi}}_1^{\dag}+m_1^2\tilde{\phi}^{\dag}_1+\epsilon\mu^2\tilde{\phi}^{\dag}_2-\frac{\lambda}{2}\tilde{\phi}_1^{\dag2}\phi_1=0\,,\\
    -\ddot{\tilde{\phi}}^{\dag}_2-3H\dot{\tilde{\phi}}_2^{\dag}-m_2^2\tilde{\phi}_2^{\dag}-\mu^2\tilde{\phi}_1^{\dag}=0\,.
\end{align}
\end{subequations}
We also have the Friedmann equation
\begin{equation}
\label{eg:H}
    H^2= -\frac{8\pi G}{3} T_{00}\,,
\end{equation}
where
\begin{equation}
\label{eq:emt}
    T^{\nu\rho}=\partial^{\nu}\tilde{\phi}^{\dag}_a\partial^{\rho}\phi_a+\partial^{\rho}\tilde{\phi}^{\dag}_a\partial^{\nu}\phi_a-g^{\nu\rho}\mathcal{L}
\end{equation}
is the stress-energy tensor. We note the signs of the first two terms, which result from the overall sign of the Lagrangian chosen in Eq.~\eqref{eq:Lag}. As in the Hermitian case, the acceleration equation is not independent and can be derived from the Friedmann constraint and Klein-Gordon equations, and is therefore not needed for what follows.

In the  $U(1)$ symmetry-broken regime, it is convenient and illustrative to scale out the vacuum expectation values of the fields by defining $\varphi_i=\phi_i/v_i$ and $\varphi_i^{\dag}=\tilde{\phi}_i^{\dag}/\tilde{v}_i$ (with $\tilde{v}_1=v_1$ and $\tilde{v}_2=-\epsilon v_2$), and to absorb a factor of $m_2$ into the coordinates (taking $\bar{x}^{\mu}=m_2x^{\mu}$). We can then write the equations of motion in the form
\begin{subequations}
\label{eq:fullsystem}
\begin{align}
    &-\ddot{\varphi}_1-3\bar{H}\dot{\varphi}_1+\bar{m}^2\varphi_1-\epsilon\bar{\mu}^2\varphi_2-\left(\bar{m}^2-\epsilon\bar{\mu}^2\right)\varphi_1^{\dag}\varphi_1^2=0\,,\\
    &-\ddot{\varphi}_2-3\bar{H}\dot{\varphi}_2-\varphi_2+\varphi_1=0\,,\allowdisplaybreaks\\
    &\bar{H}^2=\frac{8\pi G}{3}v_1^2\Bigg[\epsilon\bar{\mu}^2\dot{\varphi}_2^{\dag}\dot{\varphi}_2-\dot{\varphi}^{\dag}_1\dot{\varphi}_1+\bar{m}^2|\varphi_1|^2+\epsilon \bar{\mu}^2|\varphi_2|^2\nonumber\\&\phantom{\bar{H}^2=} -\epsilon \bar{\mu}^2\left(\varphi_1^{\dag}\varphi_2+\varphi_2^{\dag}\varphi_1\right)-\frac{1}{2}\left(\bar{m}^2-\epsilon\bar{\mu}^2\right)|\varphi_1|^4\Bigg]\,,
\end{align}
\end{subequations}
along with the Hermitian-conjugate expressions. Herein, we have defined
\begin{equation}
\bar{H}=\frac{H}{m_2}\,,\qquad \bar{m}=\frac{m_1}{m_2}\qquad \text{and}\qquad \bar{\mu}=\frac{\mu^2}{m_2^2}\,.
\end{equation}
In these variables, the broken regime of the global $U(1)$ symmetry occurs for $\bar{m}^2>\epsilon\bar{\mu}^2$ and the (Minkowskian) $\mathcal{PT}$-broken regime of the non-Hermitian system ($\epsilon=+1$) then corresponds to $\bar{\mu}^2>1$, in which case the larger kinetic term is that of the field $\varphi_2$, which contributes positively to the energy density due to our earlier choice for the overall sign of the Lagrangian. 

We now proceed to isolate the dynamics of the radial and azimuthal components of the fields, writing $\varphi_a=R_ae^{i\theta_a}$, and separating the real and imaginary parts of the Klein-Gordon equations. This yields the following system of equations:
\begin{subequations}
\label{eqs:eom}
\begin{align}
&-\ddot{R}_1+R_1\dot{\theta}_1^2-3\bar{H} \dot{R}_1+\bar{m}^2R_1-\epsilon\bar{\mu}^2R_2 \cos(\theta_1-\theta_2)\nonumber\\&\qquad-\left(\bar{m}^2-\epsilon\bar{\mu}^2\right)R_1^3=0\,,\\
&-\ddot{R}_2+R_2\dot{\theta}_2^2-3\bar{H} \dot{R}_2\nonumber\\&\qquad-R_2+R_1\cos(\theta_1-\theta_2)=0\,,\\
&-R_1\ddot{\theta}_1-2\dot{R}_1\dot{\theta}_1-3\bar{H}R_1\dot{\theta}_1\nonumber\\&\qquad+\epsilon \bar{\mu}^2R_2\sin(\theta_1-\theta_2)=0\;,\\
&-R_2\ddot{\theta}_2-2\dot{R}_2\dot{\theta}_2-3\bar{H}R_2\dot{\theta}_2\nonumber\\&\qquad+R_1\sin(\theta_1-\theta_2)=0\,,\allowdisplaybreaks\\
&\bar{H}^2=\frac{8\pi G}{3}v_1^2\Bigg[\epsilon\bar{\mu}^2\left(\dot{R}_2^2+R_2^2\dot{\theta}_2^2\right)-\left(\dot{R}_1^2+R_1^2\dot{\theta}_1^2\right)\nonumber\\&\phantom{\bar{H}^2=}+\bar{m}^2R_1^2+\epsilon \bar{\mu}^2R_2^2-2\epsilon \bar{\mu}^2R_1R_2\cos(\theta_1-\theta_2)\nonumber\\&\phantom{\bar{H}^2=}-\frac{1}{2}\left(\bar{m}^2-\epsilon\bar{\mu}^2\right)R_1^4\Bigg]\,.
\end{align}
\end{subequations}

We will now assume that there exists a steady state dynamics at late times with $
\bar{H}$ constant. (We will confirm numerically that this is indeed the case below.) We can then take $\ddot{\theta}_j$, $\ddot{R}_j$ and $\dot{R}_j$ to zero, and we obtain a system of coupled first-order ODEs
\begin{subequations}
\label{eqs:eom2}
\begin{align}
&R_1\dot{\theta}_1^2+\bar{m}^2R_1-\epsilon\bar{\mu}^2R_2 \cos(\theta_1-\theta_2)\nonumber\\&\qquad-\left(\bar{m}^2-\epsilon\bar{\mu}^2\right)R_1^3=0\,,\label{eq:eom2a}\\
&R_2\dot{\theta}_2^2-R_2+R_1\cos(\theta_1-\theta_2)=0\,,\label{eq:eom2b}\\
&-3\bar{H}R_1\dot{\theta}_1+\epsilon \bar{\mu}^2R_2\sin(\theta_1-\theta_2)=0\,, \label{eq:eom2c}\\
&-3\bar{H}R_2\dot{\theta}_2+R_1\sin(\theta_1-\theta_2)=0\,. \label{eq:eom2d}
\end{align}
\end{subequations}
In the Hermitian case ($\epsilon = -1$), from Eqs.~\eqref{eq:eom2c}--\eqref{eq:eom2d}, we find that $\theta_1 \approx \theta_2$, $\dot{\theta}_1 \approx \dot{\theta}_2 \approx 0$, and we can then use Eqs.~\eqref{eq:eom2a}--\eqref{eq:eom2b} to obtain $R_1 \approx R_2\approx 1$. The resulting Hubble rate is given by $\bar{H}^2=(4\pi G/3)v_1^2(\bar{m}^2+\bar{\mu}^2)$, which is positive  because $T_{00}$ is negative at the minimum of the potential (where the value of the potential is negative) due to the wrong overall sign choice on the Lagrangian~\eqref{eq:Lag} compared to a standard Hermitian theory.

More generally (keeping $\epsilon$ a free parameter), it follows that the difference between the phases $\delta \theta=\theta_1-\theta_2$ evolves according to
\begin{equation}
    \label{eq:deltathetadot}
    -3\bar{H}\dot{\delta \theta}+\left(\epsilon\bar{\mu}^2\frac{R_2}{R_1}-\frac{R_1}{R_2}\right)\sin\delta \theta=0\,.
\end{equation}
For $\bar{H}=\text{const.}$, the solution is
\begin{equation}
    \delta \theta=2{\rm arccot}\left\{A\exp\left[\left(\frac{R_1}{R_2}-\epsilon \bar{\mu}^2\frac{R_2}{R_1}\right)\frac{\bar{t}}{3\bar{H}}\right]\right\}\,,
\end{equation}
where $A$ is a constant. In the limit $\bar{t}=m_2t\to \infty$, we therefore have that
\begin{equation}
    \label{eq:limits}
    \lim_{m_2 t\to \infty}\delta\theta=\begin{cases} 0\,,&\qquad \epsilon\mu^4/m_2^4\frac{R_2^2}{R_1^2}<1\\ \text{const.}\,,&\qquad \epsilon\mu^4/m_2^4\frac{R_2^2}{R_1^2}\geq 1\end{cases}\,.
\end{equation}
Notice that the lower case can arise only for the non-Hermitian theory (when $\epsilon=+1$) above the exceptional point identified earlier, i.e., in the would-be $\mathcal{PT}$-broken regime in Minkowski spacetime. We remark that the sign of the constant is unimportant, given the sinusoidal nature of the potential for the phases and therefore amounts to an overall relative phase shift.

Proceeding further with the non-Hermitian case, the above expectation that $\delta\theta=\theta_1-\theta_2 = \rm{const.}$ at late times implies that $ \dot{\theta}_1 \approx \dot{\theta}_2 \equiv \dot{\theta}$.  We can then use Eqs.~\eqref{eq:eom2c}--\eqref{eq:eom2d} to obtain
\begin{equation}
R_1=\bar{\mu}R_2 \,.
\label{eq:R1R2}
\end{equation}
Note that, as $R_1$ and $R_2$ are real, this solution is only valid in the non-Hermitian case ($\epsilon=+1$). Making use of Eq.~\eqref{eq:R1R2} in Eq.~\eqref{eq:deltathetadot} gives a self-consistent prediction of $\dot{\delta\theta}=0$ and $\delta\theta=2{\rm arccot}A=\rm{const.}$ at late times. Moreover, returning to Eq.~\eqref{eq:limits}, we have $\mu^4R_2^2/(m_2^4R_1^2)=1$, such that $\delta\theta=\text{const.}$ is a generic feature of the non-Hermitian case in the $\mathcal{PT}$-broken regime.

\begin{table}[t!]
\begin{tabular}{c|c||c|c}
Parameter & Value & Initial condition & Value\\
\hline\hline
$\bar{m}$ & $4$ & $R_{1,2}(0)$ & $1$\\
$\bar{\mu}$ & $1.5^2 > 1$ & $\dot{R}_{1,2}(0)$ & $0.5$\\
$G v_1^2$ & $1/64$ & $\theta_{1,2}(0)$ & $\pi$\\
$\bar{m}^2-\bar{\mu}^2$& $>1$ & $\dot{\theta}_{1,2}$ & $0.1$\\
\end{tabular}
\caption{Parameters and initial conditions for the illustrative numerical run shown in Fig.~\ref{fig:nh}. Note that the chosen values of $\bar{m}$ and $\bar{\mu}$ place this in the broken regimes of the $\mathcal{PT}$ and global $U(1)$ symmetries. We work in units of $m_2=1$.}
\label{tab:params}
\end{table}

Continuing, the equations of motion~\eqref{eqs:eom2} simplify to
\begin{subequations}
\label{eqs:eom3}
\begin{gather}
\dot{\theta}^2+\bar{m}^2-\bar{\mu} \cos\delta\theta-\left(\bar{m}^2-\bar{\mu}^2\right)R_1^2=0\,,\label{eq:eom3a}\\
\dot{\theta}^2-1+\bar{\mu}\cos\delta\theta=0\,,\label{eq:eom3b}\\
-3\bar{H}\dot{\theta}+\bar{\mu}\sin\delta\theta=0\,. \label{eq:eom3c}
\end{gather}
\end{subequations}
Taken with the late-time limit of the Friedmann equation
\begin{align}
\label{eq:Fr}
\bar{H}^2 &= \frac{8\pi G}{3}v_1^2\Bigg[\left(1+\bar{m}^2\right)R_1^2-2\bar{\mu}R_1^2\cos\delta\theta\nonumber\\&\phantom{=}-\frac{1}{2}\left(\bar{m}^2-\bar{\mu}^2\right)R_1^4\Bigg]\,,
\end{align}
we therefore have a system of four equations in four unknowns ($R_1,\dot{\theta},\delta\theta, \bar{H}$).  This system can be solved analytically. However, the equation for $R_1^2$ is a cubic, and the result is therefore not particularly enlightening, and we do not present it here. Even so, using Eqs.~\eqref{eq:eom3b}--\eqref{eq:eom3c}, we find that $\bar{H}$ and $\dot{\theta}$ are related by
\begin{equation}
\label{eq:Hthetadot}
\bar{\mu}^2=1+\dot{\theta}^4-2\dot{\theta}^2+9\bar{H}^2\dot{\theta}^2 \,,
\end{equation}
and the Hubble rate is uniquely fixed by the late-time relative phase $\delta\theta$ via
\begin{equation}
    \label{eq:Hdeltatheta}
    \bar{H}^2=\frac{\bar{\mu}^2}{9}\frac{\sin^2\delta \theta}{1-\bar{\mu}\cos\delta\theta}\,,
\end{equation}
independent of the initial conditions (notwithstanding that those initial conditions must be consistent with reaching this late-time attractor). The dependence on the gravitational coupling enters through the implicit dependence of $\delta\theta$ on $R_1$ as fixed by Eq.~\eqref{eq:eom3a} and the late-time limit of the Friedmann equation~\eqref{eq:Fr}.

\begin{figure*}[t!]
\centering
\subfloat[][Minimum value of the potential unshifted.]{\shortstack{
\hspace{0.5em}\includegraphics[scale=0.6]{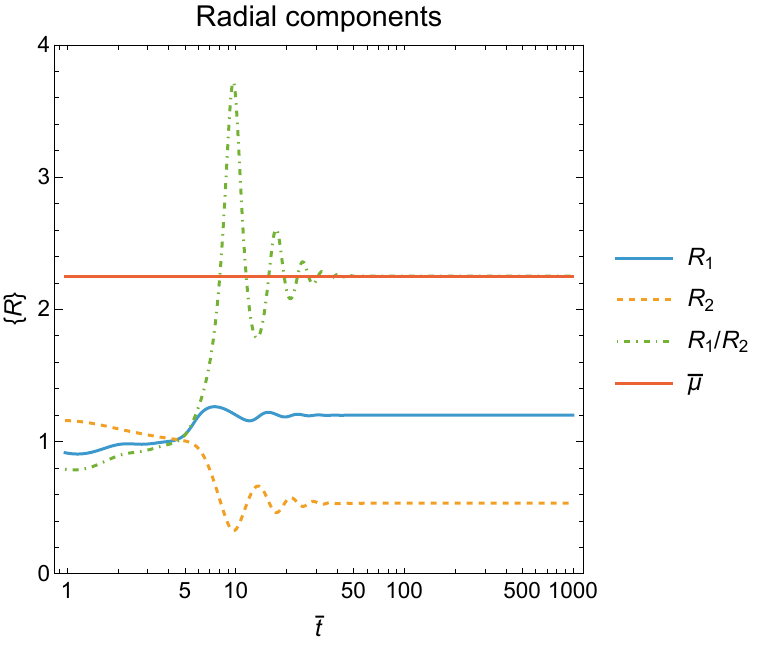}\\\vspace{1em}
\hspace{0.5em}\includegraphics[scale=0.6]{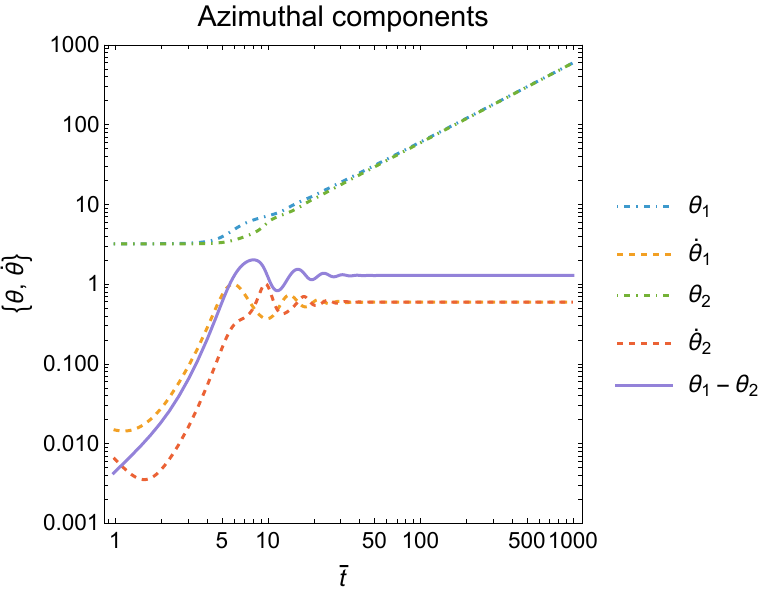}\\\vspace{1em}
\hspace{0.5em}\includegraphics[scale=0.6]{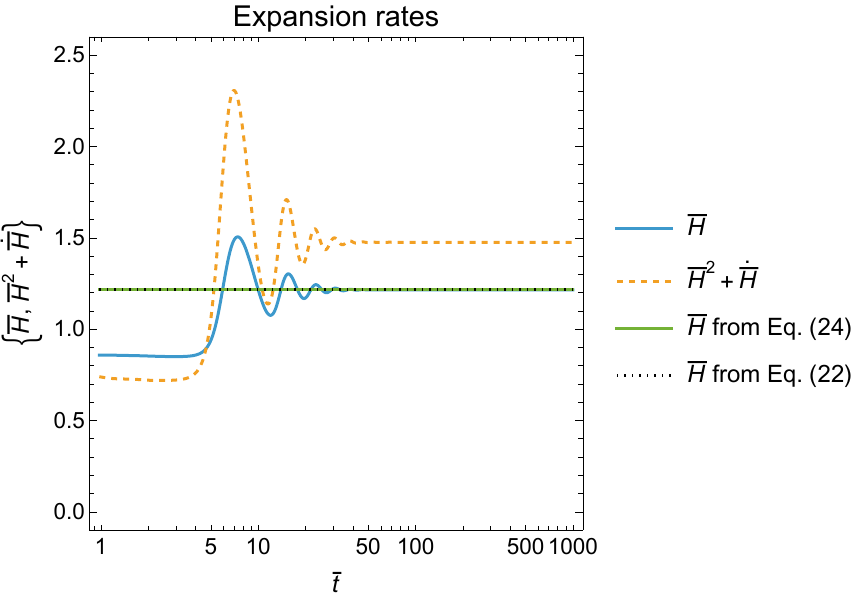}}}
\subfloat[][Minimum value of the potential shifted to zero.]{\shortstack{
\hspace{0.5em}\includegraphics[scale=0.6]{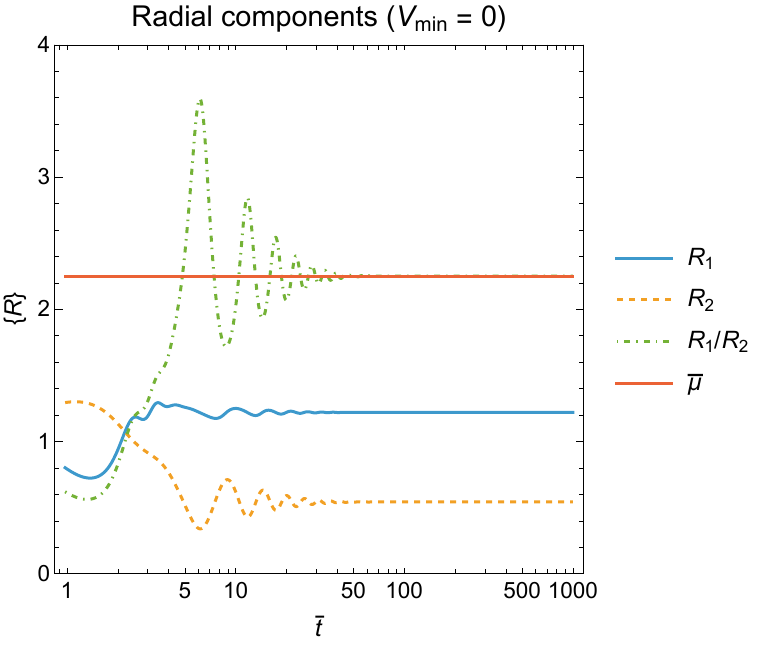}\\\vspace{1em}
\hspace{0.5em}\includegraphics[scale=0.6]{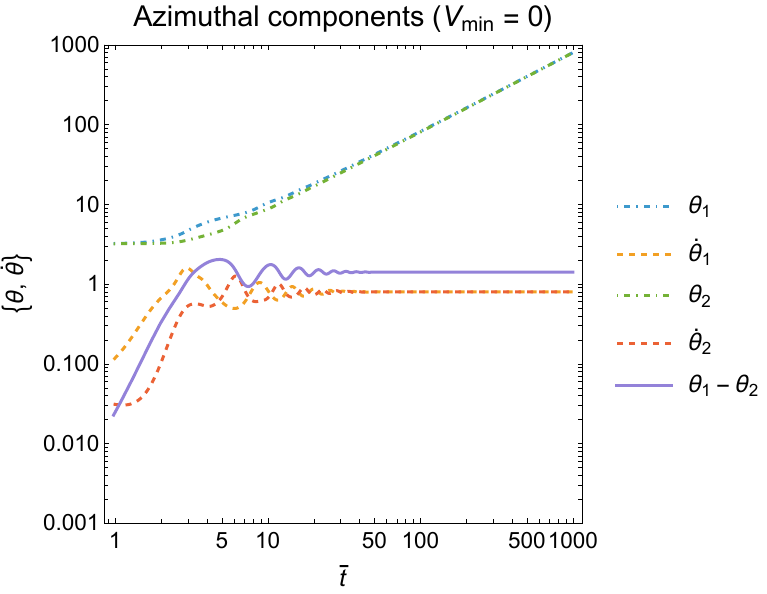}\\\vspace{1em}
\hspace{0.5em}\includegraphics[scale=0.6]{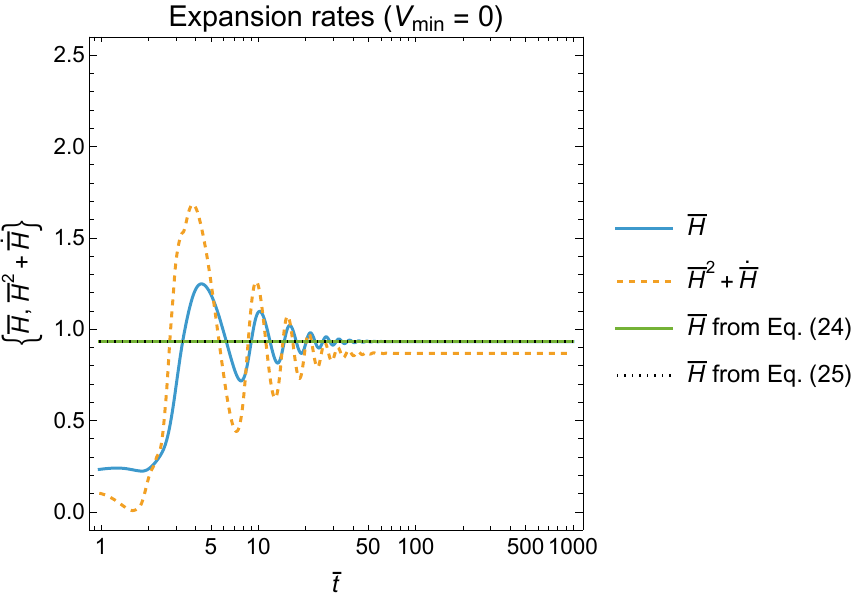}}
}

\caption{Time evolution of the radial (top) and azimuthal components (middle), and the Hubble rate and acceleration (bottom) for the non-Hermitian case. The potential has been shifted in (b) such that its value at the minima is zero. The parameters and initial conditions are as given in Tab.~\ref{tab:params}. We see that the radial components settle to fixed values away from the minimum, with the fixed ratio given by Eq.~\eqref{eq:R1R2} and that the azimuthal components reach a constant rate of roll and constant relative phase, providing the constant Hubble rate that matches the analytic expectations from Eqs.~\eqref{eq:Hdeltatheta} and the Friedmann constraint.}
\label{fig:nh}
\end{figure*}

One might argue that, as in the Hermitian case above, the late-time expansion is simply being supported by the non-vanishing value of the potential at the minimum. However, subtracting this only shifts the Hubble rate~\eqref{eq:Fr} by $-\frac{4\pi G}{3}v_1^2(\bar{m}^2-\bar{\mu}^2)$, and the result
\begin{align}
\label{eq:Frshift}
\bar{H}^2 &= \frac{8\pi G}{3}v_1^2\Bigg[\left(1+\bar{m}^2\right)R_1^2-2\bar{\mu}R_1^2\cos\delta\theta\nonumber\\&\phantom{=}-\frac{1}{2}\left(\bar{m}^2-\bar{\mu}^2\right)(1+R_1^4)\Bigg]
\end{align}
remains non-vanishing. Noting that Eqs.~\eqref{eq:Fr} and~\eqref{eq:Frshift} are derived using $R_1=\bar{\mu}R_2$, we see that this is because the radial components of the fields do not settle to the minimum at late times due to the centripetal acceleration provided by the azimuthal components of the field, which roll \emph{ad infinitum} at a constant rate at late times. In this sense, the behaviour of the radial components is akin to perfect slow-roll, with these components pinned to the slope of the potential by the motion of the azimuthal components. We mention in passing that related behaviour of a constant radial component and constant angular velocity can be seen in a class of two-field inflationary attractor models, known as ‘shift-symmetric orbital inflation’, whose behaviour is strongly multi-field but whose predictions are remarkably close to those of single-field inflation \cite{Achucarro:2019pux}.

The precise behaviour described above is illustrated in Fig.~\ref{fig:nh}, where the system of equations~\eqref{eqs:eom} has been solved numerically using Mathematica's native \texttt{NDSolve} routine for the parameters and initial conditions provided in Tab.~\ref{tab:params}. Moreover, the value of the potential at the minimum has been shifted to zero. We note that, having shifted the minimum of the potential, it is necessary to choose initial conditions that are compatible with a negative value of $T_{00}$. The expected analytical scalings are confirmed --- namely Eqs.~\eqref{eq:Hthetadot}, \eqref{eq:R1R2} and~\eqref{eq:Hdeltatheta} --- and these are plotted in Fig.~\ref{fig:nh} for comparison.

Interestingly, by inspection of the equations of motion with $\delta\theta\to{\rm const.}$, the effective late-time potential for each $\theta_a$ is linear in $\theta_a$, while the azimuthal direction itself is, of course, compact. The dynamics is reminiscent of M.~C.~Escher's lithograph ``Ascending and Descending'', with the azimuthal components rolling forever down the infinite staircase (driven by the inherent instability of the $\mathcal{PT}$-broken regime) but unable to accelerate due to the (self-generated) Hubble friction.

\begin{figure}[b!]
\includegraphics[scale=0.6]{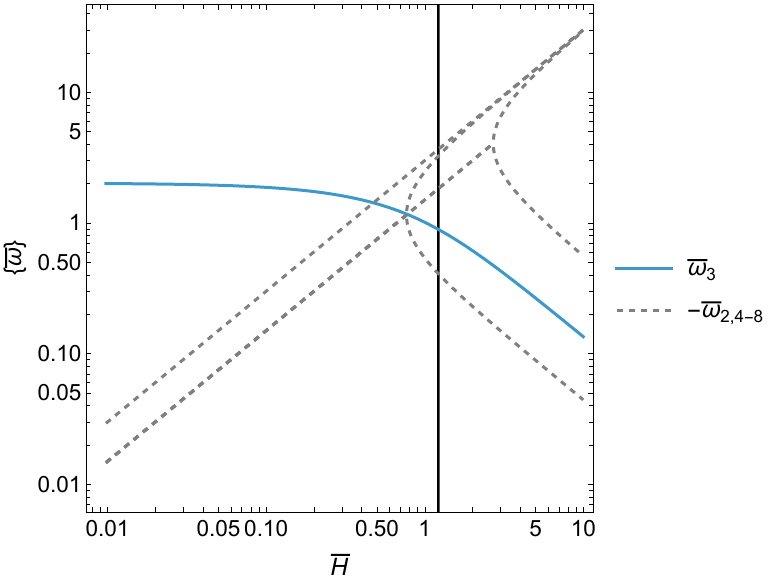} 
\caption{Imaginary parts of the eigenfrequencies of fluctuations around the minimum of the potential as a function of $\bar{H}$ for the parameter choices of Fig.~\ref{fig:nh}. The frequency with positive imaginary part is shown with a solid blue line, whereas all other frequencies have a negative imaginary part and are shown with dashed grey lines. We have chosen not to identify the latter modes individually, since they merge at the various exceptional points visible in the plot. The vertical line is the late-time value of $\bar{H}$ from Fig.~\ref{fig:nh} (a).}
\label{fig:eigens}
\end{figure}

The behaviour of the non-Hermitian system can also be understood in terms of the eigenspectrum of fluctuations about the minimum of the potential. We consider the equations of motion~\eqref{eqs:eom} for small perturbations around constant background fields, writing
\begin{equation}
R_a = \bar{R}_a+\delta R_a\,, \quad \theta_a = \bar{\Theta}t+\bar{\theta}_a+\delta \theta_a\,,
\end{equation}
and take $\bar{R}_1=\bar{R}_2=1$. The background equations then yield $\bar{\Theta}=0$, $\bar{\theta}_1=\bar{\theta}_2$, and $\bar{H}$ constant. Working to first order in the perturbations $\delta R_a$ and $\delta\theta_a$, we find that the normalised eigenfrequencies $\bar{\omega}\equiv \omega/m_2$ are determined from 
\begin{widetext}
\begin{equation}
    {\rm det}\begin{pmatrix} \bar{\omega}^2+3 i \bar{H} \bar{\omega} +3 \epsilon \bar{\mu}^2 - 2 \bar{m}^2 & -\epsilon \bar{\mu}^2 & 0 & 0 \\ 1 & \bar{\omega}^2+3i \bar{H} \bar{\omega}-1 & 0 & 0\\
    0 & 0 & \bar{\omega}^2+3 i \bar{H} \bar{\omega}+\epsilon\bar{\mu}^2 & -\epsilon \bar{\mu}^2\\ 0 & 0 & 1 & \bar{\omega}^2+3i \bar{H} \bar{\omega}-1\end{pmatrix}=0\,,
\end{equation}
This equation admits 8 solutions, and we obtain
\begin{subequations}
\label{eq:eigenvalues}
\begin{gather}
\bar{\omega}_1 = 0\, , \qquad
\bar{\omega}_2 = -3i \bar{H}\, \qquad
\bar{\omega}_{3,4} = - \frac{3i \bar{H}}{2} \pm \frac{i}{2} \sqrt{9 \bar{H}^2+4 \epsilon \bar{\mu}^2-4}\, \\
\bar{\omega}_{5-8}  = - \frac{3 i\bar{H}}{2} \pm\frac{i}{2} \sqrt{9 \bar{H}^2+6 \epsilon \bar{\mu}^2-4 \bar{m}^2 \pm 2 i \sqrt{-9 \bar{\mu}^4-2 \epsilon\bar{\mu}^2-4 \bar{m}^4+12 \epsilon \bar{\mu}^2 \bar{m}^2+4 \bar{m}^2-1}-2} \,.
\end{gather}
\end{subequations}
\end{widetext}
The behaviour of the spectrum is different in the Hermitian and non-Hermitian cases. In the Hermitian scenario ($\epsilon=-1$), the imaginary part of these energies is always negative, and determined only by the Hubble damping. On the other hand, for the non-Hermitian system, two of the modes can obtain positive imaginary parts, giving rise to growing modes. The eigenfrequencies for the parameter region shown in Fig.~\ref{fig:nh} are plotted in Fig.~\ref{fig:eigens} as a function of $\bar{H}$, wherein we see that there exists one mode with a positive imaginary part, driving the system away from the minimum of the potential.

Thus, in this letter, we have arrived at a novel mechanism for driving accelerated cosmological expansion, which exploits the $\mathcal{PT}$-broken regime of a pseudo-Hermitian field theory. Further work is, of course, warranted, particularly in relation to the pseudo-Hermitian \emph{quantum} dynamics of this theory. Additionally, one might expect unique signatures, e.g., in the power spectra, where the necessary pseudo-scalar nature of one field will also have an impact. One might also wonder whether the degree of non-Hermiticity can itself be made dynamical to control the onset and end of periods of accelerated expansion, either for inflation or dark energy.  We leave such considerations and parameter scans for this and similar models for future work. Even so, the present application serves to show that genuinely novel phenomenology can be achieved within the framework of pseudo-Hermitian (quantum) field theory and the $\mathcal{PT}$-broken regime can be physically relevant once we move to dynamical spacetimes.

\begin{acknowledgements}
This work was supported by the University of Manchester [ES], a Nottingham Research Fellowship from the University of Nottingham [PM], the Science and Technology Facilities Council (STFC) [Grant No.\ ST/X00077X/1] [PM] and [Grant No.\ ST/X000672/1] [EJC], and a United Kingdom Research and Innovation (UKRI) Future Leaders Fellowship [Grant Nos.\ MR/V021974/1 and\ MR/V021974/2] [AL \& PM].

No data were created or analysed in this study.
\end{acknowledgements}


\begin{thebibliography}{99}
\bibitem{Endnote2}
Since we deal here with only $c$-number fields, the transposition implicit in the definition of the Hermitian conjugate is, of course, redundant.
\end{thebibliography}
\end{document}